\documentclass[twocolumn,aps,prl,floatfix]{revtex4}
\usepackage{graphicx,graphics,color,epsfig}
\usepackage{amsmath}
\usepackage{amssymb}

\begin{document}

\title{Window measurements of simulations in random systems}
\author{Tota Nakamura$^1$ and Takayuki Shirakura$^2$}
\affiliation{$^1$College of Engineering, Shibaura Institute of Technology, Saitama 337-8570, Japan}
\affiliation{$^2$Faculty of Humanities and Social Science, Iwate University,
$^2$Faculty of Humanities and Social Science, Iwate University,
Morioka 020-8550, Japan}

\date{\today}

\begin{abstract}
Numerical studies in random systems are plagued with
strong finite-size effects and boundary effects.
We introduce a window-measurement method as a practical solution
to these difficulties.
We observe physical quantities only within a subsystem
located in the midst of a whole system and scale them with
the correlation length estimated in the subsystem.
Both equilibrium data and nonequilibrium data with different 
system sizes and different window sizes fall onto a single scaling function.
It suggests that the correction-to-scaling terms become very small.
We confirm the validity in 
the $\pm J$ Heisenberg spin glass model in three dimensions.
The spin-glass and chiral-glass transition temperatures are estimated 
to be very close to each other.
\end{abstract}

\maketitle

{\it Introduction}-
Numerical studies in condensed matter physics have made remarkable progresses
in accordance with developments in high-performance computing.\cite{MCreview}
It is now possible to observe a precise logarithmic correction term of the
Kosterlitz-Thouless transition in a two-dimensional XY model.\cite{komura}
The thermodynamic limit may be almost at hand in uniform systems.
The situation is quite different in random systems.
Generally, we need to take an average over many random samples.
It takes a very long time to equilibrate a system
in Monte Carlo (MC) simulations.
These shortcomings restrict us to treat relatively 
smaller system sizes compared to the uniform systems.
However,
it should be noted that
we need to treat larger lattice sizes in random systems
because they possess complex internal structures.
Therefore, we are staying much before the thermodynamic limit
in a numerical study of random systems.

We explain the situation using an example of 
spin glasses (SG),\cite{SGReview3,KawashimaRieger}
which are disordered magnets 
characterized by frustration and randomness.
Spin glasses are regarded as a prototype of many complex systems.
Efficient numerical algorithms that can overcome difficulties in 
SG simulations have been successfully applied to many complex systems.
The temperature-exchange algorithm\cite{hukutempexchange}
is a typical example.
Although this efficient algorithm is applied,
a linear system size, $L$,  equilibrated in three-dimensional
spin glass systems is restricted to $L=48$.\cite{fernandez,viet2,viet}
Then, 
strong finite-size effects appear in the simulation data.
Determinations of the phase transition temperature and critical exponents
may be influenced by the way how the size effects are treated.
In the three-dimensional Heisenberg SG model, 
the ``spin-chirality coupling or decoupling'' problem is still under debate.%
\cite{Olive,KawamuraH1,HukushimaH,matsubara1,matsubara2,matsubara3,%
matsumoto-huku-taka,nakamura,Lee,berthier-Y,picco,HukushimaH2,campos,%
Lee2,viet2,viet,fernandez}
It argues whether the spin-glass transition and the chiral-glass (CG)
transition occur simultaneously or not.
There have been reported two opposite conclusions\cite{fernandez,viet2,viet}
by treating the correction-to-scaling terms in different ways.
In the three-dimensional Ising SG model,
estimated values of the critical exponent $\nu$ vary from 1.4 to 2.7,
and the issue remains unsolved as ``big $\nu$ or small $\nu$?''.%
\cite{Bhatt,Ogielski,KawashimaY,palassini,ballesteros,maricampbell,nakamura2,%
campbell-huku-taka,hasenbusch,nakamuraxisca}
Hukushima and Campbell \cite{hukushimacampbell} noted that 
finite-size correction terms show a non-monotonic behavior
in the three-dimensional Ising SG model. 
They claimed that there exists a crossover size, $L=24$,
where the correction-to-scaling terms change the sign.
It may be necessary to perform a finite-size scaling analysis 
using only larger sizes.
It is a very hard task considering our computational environment
at present.

Why simulations on random systems encounter such severe finite-size effects?
We consider that one possible answer is the boundary conditions.
The periodic boundary(PB) conditions have been used in most simulations.
It is adopted originally in uniform systems
to retrieve the translational invariance.
However, its application to random systems is not trivial.
It produces an artificial and
unexpected symmetry: a translation of $L$ lattice spacings.
In principle, random systems must not have any translational symmetry.
Alternatively,
we may impose the open boundary(OB) conditions in random systems.
However, the coordination number differs between bulk spins and surface spins.
This has been considered to produce stronger finite-size effects from
the simulational experiences in uniform systems.
Shirakura and Matsubara \cite{shirakuraB} studied on two boundary
conditions applied to the Heisenberg spin glass model.
They observed distribution functions of the spin-glass order parameter
in the periodic system and in the open system.
Two results are quite different at low temperatures 
for system sizes up to $L=31$.
However, we cannot make a fair judgment of which boundary condition
is proper.

In this paper,
we introduce a practical solution to the difficulties mentioned above.
This is a window measurement scheme.
The basic idea is same as a window overlap.\cite{newman,marinari}
A distribution function of the spin-glass order parameter, $P(q)$, 
was observed only within a window region located in the midst of
a whole system.
It has been known that the window overlap is less influenced by the 
size volume and the boundary conditions.
We may regard the window measurement as a compromise between 
the periodic and the open boundary conditions.
Usually, we simulate a size-$L$ periodic 
system and observe physical quantities for the whole system.
In the window measurement scheme, we only observe quantities in an open
size-$B$ subsystem ($B<L$) located in the midst of the system.
It is also possible to define different window sizes and collect different
series of window data at once.
However, window measurements for extensive variables such as
the susceptibility have not been applied yet.
One reason is their complicated finite-size dependences
both on $L$ and $B$.
We solve this difficulty by the correlation-length scaling.\cite{nakamuraxisca}

{\it Model and simulation details}-
We study the $\pm J$ Heisenberg SG model on a simple cubic lattice:
\begin{equation}
   {\cal H} = - \sum_{\langle i,j \rangle}
      J_{ij} S_i  S_j.
\label{eq:H1}
\end{equation}
The summation runs over all the nearest-neighbor spin pairs.
The interactions $J_{ij}$ take two values, $\pm J$,
with the same probability. The temperature $T$ is scaled by $J$. 
The lattice is of the form $N = L \times L \times (L+1)$, and
$L$ is an odd number.
We calculate the SG and CG susceptibility, $\chi_\mathrm{s}$ 
and $\chi_\mathrm{c}$, 
and 
the SG and CG correlation length, $\xi_\mathrm{s}$ and $\xi_\mathrm{c}$.
We perform both static(equilibrium) simulations and 
dynamic(nonequilibrium) simulations.
A definition for the correlation length 
in the static simulations
is the Ornstein-Zernike formula.\cite{ornstein}
That in the dynamic simulations is a modified version 
of the Ornstein-Zernike formula proposed previously
for the dynamic correlation length.\cite{nakamuraxisca}

In the static simulations, we focus on the difference between open and 
periodic boundary conditions, and
observe how the difference vanishes as the system/window size increases.
One MC step consists of one heat-bath update, 
one temperature-exchange update, and $L/2$ overrelaxation updates.
We present data of the linear size $L=15, 31$, and 41.
Numbers of MC step are $54000$($L=15$), and $1.8\times 10^6$($L=31$ and 41).
Sample numbers are 128($L=15$), 96($L=31$), and 16($L=41$).
SG and CG order parameters are evaluated using an overlap between
two real replicas.

In the dynamic simulations, 
relaxation functions of the susceptibility and the correlation length
are studied.
We present data of the liner size $L=19, 39, 79$, and 159.
One MC step consists of one heat-bath update,
1/20 Metropolis update(once in every 20 steps),
and 124 overrelaxation updates.
Sample numbers are 900($L=19$), 211($L=39$), 65($L=79$), and 20($L=159$). 
SG and CG order parameters are evaluated using 435 overlaps among
thirty real replicas.
We apply only the skew-periodic boundary condition.
Numerical error bars are estimated in regard to the sample average.

\begin{figure}
  \resizebox{0.48\textwidth}{!}{\includegraphics{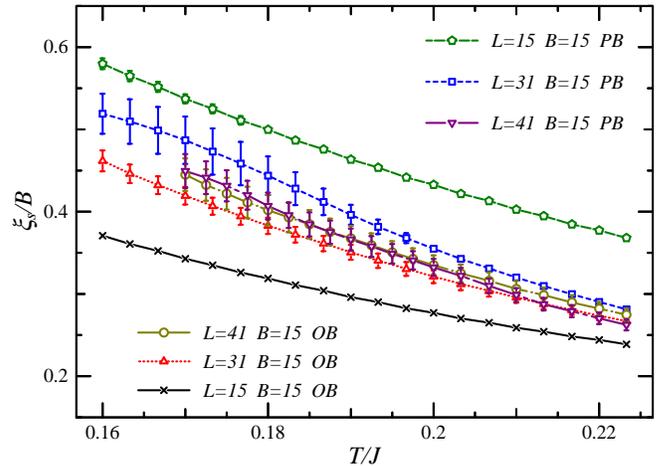}}
  \caption{
(Color online)
Results of window measurements in static simulations.
The SG correlation-length ratio, $\xi_{\rm s}/B$, is
plotted against the temperature.
$B$ is a size of window.
PB and OB stand for periodic boundary and open boundary conditions.
}
\label{fig:staticB}
\end{figure}

{\it Results}-
First, we show the static simulation data.
The equilibrium SG correlation-length ratio, $\xi_\mathrm{s}/B$, is plotted
against the temperature 
when $B=15$ in Fig.~\ref{fig:staticB}.
The equilibrium values depend on the boundary conditions when $B=L$.
As $L$ increases,
the window-measurement data of PB and OB approach each other.
The data of $L=41$ are independent of the boundary conditions.
We obtain similar results for the equilibrium CG correlation length
ratio and the distribution function of order parameters. 
We can neglect the boundary effects if we set $B/L < 1/3$.
So, what we have to do is the following procedures:
(i) set the window ratio as $B/L=1/3$;
(ii) collect equilibrium data for different $B$;
(iii) perform the finite-$B$ scaling analyses.
However, this procedure does not sound realistic because the equilibrium
simulation up to now is restricted to $L=48$, which gives the upper
bound for $B$ is 16.
Therefore, we adopt a dynamic scaling 
approach\cite{NerReview,ozekiito-SG,nakamura5,yamamoto1,nakamura4}
that can handle larger lattice sizes.
We also try to relax the ratio requirement, $B/L<1/3$, which discards
most of the simulated spins.

\begin{figure}
\begin{center}
  \resizebox{0.40\textwidth}{!}{\includegraphics{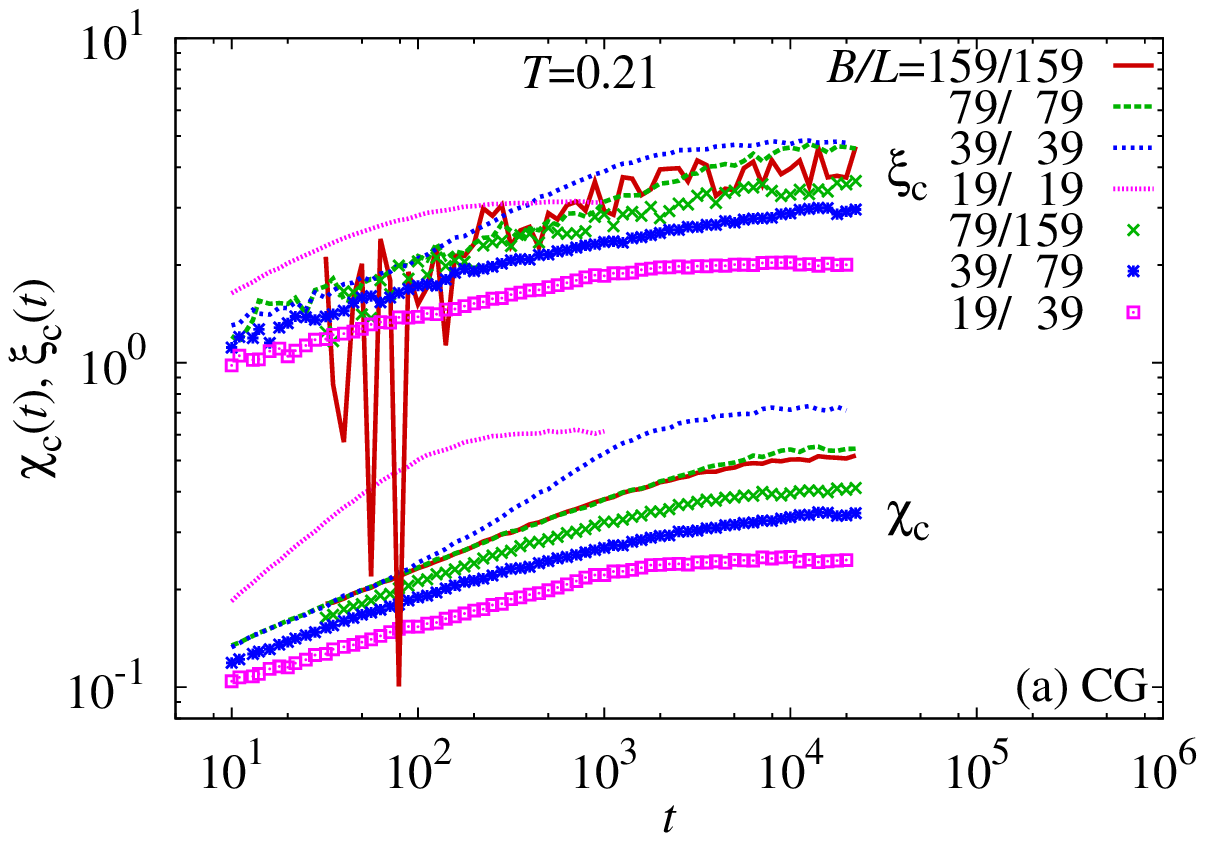}}
  \resizebox{0.40\textwidth}{!}{\includegraphics{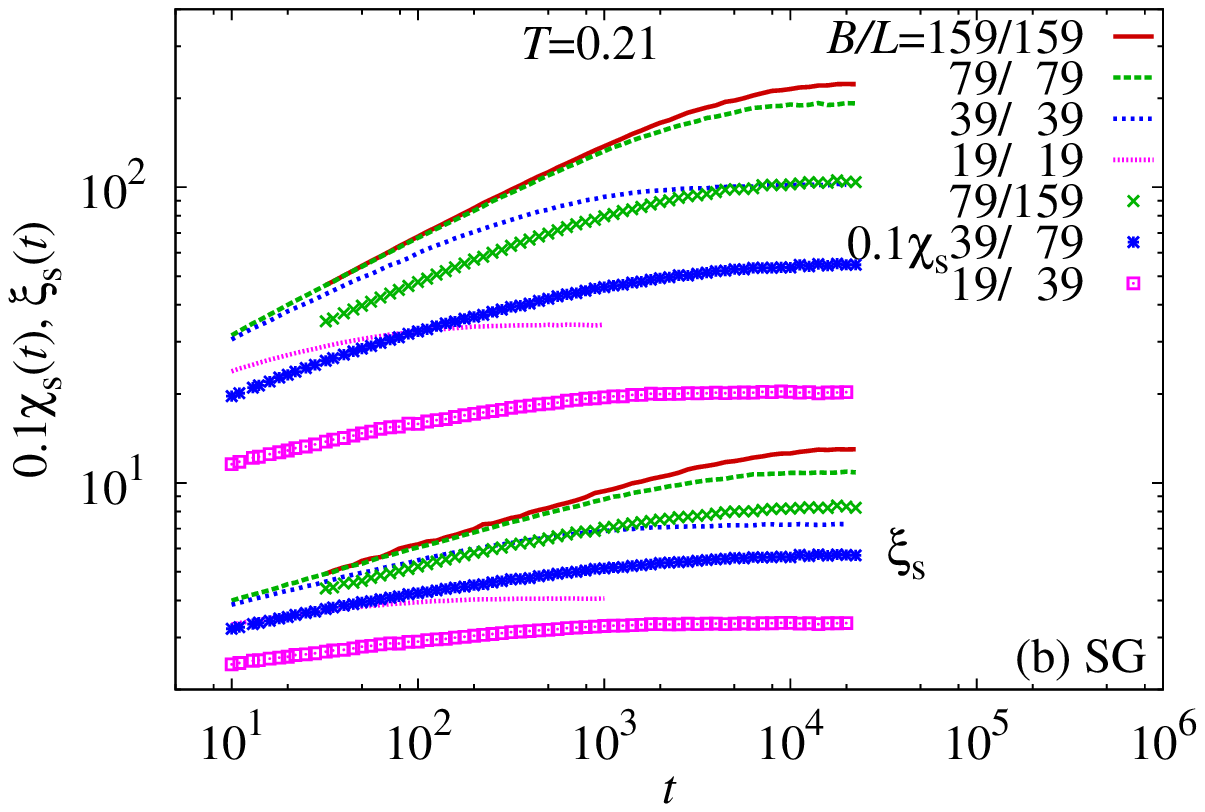}}
\end{center}
  \caption{
(Color online)
Relaxation functions of the susceptibility and the correlation length 
for whole measurements (lines) and window measurements (symbols).
Error bars are smaller than line width for SG data 
but they are in a same order as data fluctuation for CG data.
}
\label{fig:sizecrossover1}
\end{figure}

Figure ~\ref{fig:sizecrossover1} shows a dynamic simulation result.
We plot relaxation functions of (a) $\chi_{\rm c}$ and $\xi_{\rm c}$,
and (b) $\chi_{\rm s}$ and $\xi_{\rm s}$.
The temperature, $T=0.21$, is considered as in a paramagnetic phase but close 
to the transition temperature.\cite{nakamura,HukushimaH2}
As was observed in the Ising SG model,\cite{hukushimacampbell}
we find a size-crossover effect in relaxation data of $\chi_\mathrm{c}(t)$
when $B=L$,
while there is no size crossover when $B<L$.
A converging value takes a maximum when $B=L=39$ and it decreases as $L$
increases or decreases.
They also deviate to the upper side when the finite-size effects appear.
It suggests that the finite-size scaling analyses using data of $L$
smaller than 39 should be carefully performed.
It is also noted that
an amplitude of each relaxation function depends on $B$, $L$, and $B/L$.
If we fix $B$ and compare 
the window-measurement results and the whole-measurement results
(lines and symbols with same color in Fig.~\ref{fig:sizecrossover1}),
we find that the window results always take smaller values 
than the whole results. 
Both results for $\chi_{\rm c}$ approach each other as $B$ increases
but  those for $\chi_{\rm s}$ do not.
These dependences are quantitatively complicated.
This ambiguity may be a reason why the window measurement scheme
has not been applied to the scaling analyses.

We apply the dynamic-correlation-length scaling 
analysis\cite{nakamuraxisca} to the window measurements.
A relaxation function of the susceptibility is plotted against 
that of the correlation length.
It should show an algebraic divergence as $\chi\sim\xi^{2-\eta}$
at the second-order transition temperature.
It is a straight line if we plot it in a log-log scale.
It exhibits a downward-bending behavior in the paramagnetic phase
and an upward-bending behavior in the ordered phase.

\begin{figure}
  \resizebox{0.48\textwidth}{!}{\includegraphics{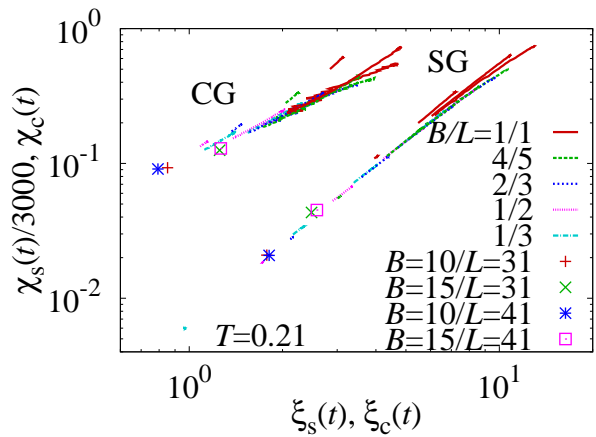}}
  \caption{
(Color online)
Relaxation functions of the susceptibility are plotted
against those of the correlation length.
Each line type corresponds to each $B/L$ ratio for
all the systems sizes, $L=$19, 39, 79 and 159(SG only).
Static simulation data of $B=10$ and 15 for $L=31$ and 41 
with OB conditions are also plotted with symbols.
Error bars are smaller than line width and symbols. 
}
\label{fig:chixi}
\end{figure}

Figure~\ref{fig:chixi} shows
finite-time and finite-size relaxation data of 
$\chi_{\rm s/c}(t, L)$  plotted against $\xi_{\rm s/c}(t, L)$. 
We also plot equilibrium data with OB conditions.
Here, the CG results of $L=159$ are omitted because the $\xi_{\rm c}$ data
include large numerical fluctuation 
as shown in Fig.~\ref{fig:sizecrossover1}(a).
In this figure
we find that all the SG window-measurement data
(nonequilibrium data of $B/L=4/5$--$1/3$ for $L=19$--$159$ and 
equilibrium data of $B=10$ and 15 for $L=31$ and 41)
ride on a single scaling function.
Equilibrium data of small sizes appear on the way of 
nonequilibrium relaxation data of large sizes.
It can be considered that the SG susceptibility increases
in accordance with the SG correlation length in a uniform manner.
Size and time, which are man-made parameters, do not matter much.
This scaling function is expected to continue smoothly 
to the thermodynamic limit.
In other words, the correction-to-scaling terms are considered as very small.
This is an advantage of the window measurement and 
the correlation-length scaling.
As a result,
the $B/L$ ratios can be set larger than $1/3$ up to $4/5$.
This improves the computational efficiency.
To the contrary,
the whole-measurement data ($B/L=1/1$; $L=19$--$159$)
show size dependences.
Small-size data deviate to the upper side and only the largest-size
data exhibit the downward bending.
We may need still larger lattices in the whole-measurement scheme
to observe size-independent thermodynamic behaviors.
\begin{figure}
  \resizebox{0.48\textwidth}{!}{\includegraphics{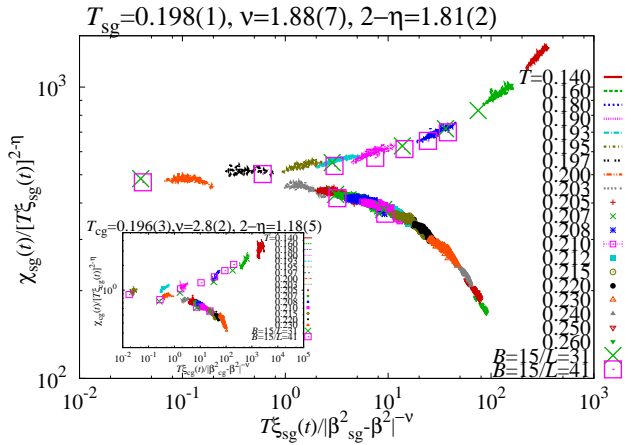}}
  \caption{
(Color online)
A dynamic scaling plot of the SG susceptibility and SG correlation lengths.
Nonequilibrium data of $B/L=4/5, 2/3, 1/2$, $1/3$ with $L=79$ and 159,
and equilibrium data of $B=15$ with $L=31$ and 41(OB conditions)
are plotted for each temperature.
The first 200 steps are discarded for nonequilibrium data.
An inset is the scaling plot for CG (up to $L=79$).
Error bars are smaller than line width and symbols.
}
\label{fig:chixisca}
\end{figure}

Since the correction-to-scaling terms are considered to be small
in the window measurements,
we can use all the finite-size and the finite-time relaxation data
of $\chi$ and $\xi$ for the dynamic scaling analysis to determine
the transition temperature and critical exponents.
Figure \ref{fig:chixisca} is the scaling plot.
The scaling parameters are estimated by the Bayesian  inference
introduced by Harada\cite{harada}.
We randomly choose 600 datasets of $(\chi(t), \xi(t))$ out of
2798 entries for $L=79$ and 159 for SG,
and performed the inference for 40 times changing the datasets and
initial values of estimated parameters.
The $\beta$-scaling method proposed by Campbell et al.\cite{campbell-huku-taka}
is also applied. 
The estimated SG transition temperature is located between 
the one obtained by Matsubara et al.\cite{matsubara3} ($T_{\rm sg}= 0.18$)
and
the one obtained by Nakamura and Endoh\cite{nakamura} ($T_{\rm sg}= 0.21$).
We plot equilibrium data of $B=15$ for $L=31$ and 41 
using the scaling parameters estimated by the nonequilibrium data.
They ride on the same scaling function very well.
It is noted that data of $B=10$ systematically deviate to the lower side.
We consider that there exists a lower bound size that can be used in the
scaling analysis.
It may be located between 10 and 15.
The scaling plot of CG is similar to that of SG.
We performed the scaling inference using only the $L=79$ data.
An estimate of $T_{\rm cg}$ is very close to $T_{\rm sg}$.
It is also consistent with the one
obtained by Hukushima and Kawamura\cite{HukushimaH2}, which gave
$T_{\rm cg}=0.194$.
Therefore, the present results suggest the simultaneous spin-chirality
transition.

{\it Summary and Discussion}-
We have proposed a window measurement scheme,
which practically solves severe finite-size and finite-time effects.
The correction-to-scaling terms sometimes cause discrepancies in
the final result of scaling analyses.
They can be made very small by scaling the window data with the
window correlation length.
We can also combine equilibrium data of small sizes and nonequilibrium
data of large sizes into one scaling analysis.
It is a great advantage in the slow-dynamic systems.

In exchange for the benefit of the window measurements,
we must accept a big waste of computations.
Almost a half of total spins are discarded even when $B/L= 0.8$
in three dimensions. 
If $B/L=1/3$, this discard ratio becomes 96\%!
However,
a recent progress in high-performance computers drastically
dropped the computational cost and allowed us to accept the waste.
We performed the present simulations by CUDA GPGPU environment
developed by Nvidia.
We could easily achieve 10 times faster simulations by a GPU with
a reasonable price (additional \$499 per node).
A rich man's algorithm that can accept the waste
may be a standard approach in computational physics from now on.

The window measurement scheme can be applied 
to incommensurate systems and long-range interacting systems, 
where finite-size effects are also severe.
It may be promising to apply it to the
two-dimensional frustrated quantum spin systems, where the treated sizes 
are also limited.\cite{kagome}
As for the Heisenberg SG problem,
we observed a simultaneous spin-chirality transition.
We also found a size crossover\cite{hukushimacampbell} 
in the CG susceptibility.

\acknowledgments
The author would like to thank Fumitaka Matsubara
for fruitful discussions and comments.
This work is supported by a Grant-in-Aid for Scientific Research 
from the Ministry of Education, Culture, Sports, Science and Technology, 
Japan (No. 21540343 and No. 22340110).

\end{document}